# A Brief Review and Perspective on the Functional Biodegradable Films for Food Packaging

Xiaoyan He[1], Yuanyuan Jin[1], Lisheng Tang[1], Ran Huang[1,2*]

[1]Department of Material Science and Engineering, Taizhou Institute of Zhejiang University, Taizhou, Zhejiang 318000, China

[2]Academy for Engineering and Applied Technology; Yiwu Research Institute; Zhuhai Fudan Innovation Institute, Fudan University, Shanghai 200433, China

*Correspondence: huangran@fudan.edu.cn

**Abstract**

High-performance, environmentally-friendly biodegradable packaging as substitutes for conventional plastics becomes severe demand to nowadays economy and society. As an aliphatic aromatic copolyester PBAT is recognized as the preferred alternative to traditional plastics. However, the relatively high cost and weak properties obstacles the widespread adoption of PBAT. Modification pertaining to improve the properties, lower the cost, and include the functional additives of PBAT is a continuous effort to meet the needs of food accessibility, antibacterial properties, oxygen resistance, high mechanical strength, stable size, low moisture absorption, and various gas permeability for commercial competitiveness.

## 1. Introduction

Plastic food packaging serves the purpose of safeguarding food quality by mitigating the risk of contamination from microbes or other substances during the many stages of manufacturing, transportation, storage, and sales. Additionally, it aids in minimizing the occurrence of food oxidation and other chemical reactions, hence prolonging the shelf life of the packaged food items. Simultaneously, it fulfills a function in the promotion of decoration, enhancement of aesthetic appeal, and augmentation of value [1,2]. Nevertheless, the majority of plastic food soft packaging materials are throwaway items that are deficient in terms of recyclability. Following their utilization, the bulk of items can only be discarded by landfilling, incineration, and several other disposal techniques. Plastic garbage, once deposited in landfills, has limited or significantly slow decomposition. Consequently, it occupies a substantial portion of land resources and contributes to the deterioration of soil, water, and other components of the environmental ecology [3-6]. The combustion of these items gives rise to noxious compounds, such as dioxins, which present substantial hazards to the health of humans. Furthermore, the significant emission of carbon dioxide resulting from incineration has contributed to the greenhouse effect and global warming. Hence, one urgent task to achieve carbon emissions reduction is to devise efficient strategies for minimizing the utilization of conventional non-biodegradable packaging materials.

Many prominent nations across the globe are expediting the development of legislation and regulations pertaining to the management of plastic pollution. They are actively endorsing and facilitating the use of high-performance, environmentally-friendly biodegradable packaging as substitutes for conventional plastics. Biodegradable packaging is a sustainable material that possesses environmentally favorable characteristics, including recyclability, degradability, and non-interference with food safety in packaging applications. The complete life cycle of biodegradable packaging materials satisfies the criteria of carbon neutrality. Following disposal, these substances can undergo reactions with microbes present in the environment, leading to their progressive decomposition under natural circumstances such as composting, anaerobic environments, and water treatment. Ultimately, this decomposition results in the formation of carbon dioxide, water, and other organic byproducts that are benign in nature. It efficiently mitigates the environmental impact associated with the carbon emissions generated throughout the life cycle of packaging materials, encompassing their manufacturing, utilization, and decomposition phases. Enhancing the utilization of biodegradable materials in the realm of food packaging is a significant approach to address the prevailing challenges associated with plastic waste management.

**2. Progress in biodegradable film packaging**

According to differences in source, biodegradable packaging materials can be divided into natural organic polymers, such as cellulose, starch, chitin/chitosan, and artificially synthesized polymers, such as polylactic acid (PLA), polybutylene adipate terephthalate (PBAT), polyhydroxyalkanoates (PHA), poly(butylene succinate) (PBS), poly(butylene succinate-co-butylene adipate) (PBSA), poly(butylene succinate-co-terephthalate) (PBST), etc. [7-10]. PBAT is a type of aliphatic aromatic copolyester that can be completely biodegradable and has good biocompatibility. It has a high elongation at break, is easy to produce and process, and can be applied to various packaging film materials and other fields. After degradation, it can enter the ecological carbon cycle. Therefore, PBAT has become a hot research topic for researchers and related enterprises in various countries in recent years [11,12]. PBAT is recognized as the preferred alternative to traditional LLDPE plastics. However, the monomer and synthesis costs of PBAT are relatively high, the price of blown film grade PBAT is generally 3-6 times that of ordinary blown film grade PE, which also affects the widespread promotion and use of PBAT in the market; Moreover, the comprehensive mechanical properties of PBAT have been improved compared to aliphatic biodegradable polyester materials such as PBS, but it is too flexible and lose the high elastic modulus of typical aliphatic polyester. The exclusive utilization of PBAT for product packaging, particularly in thin film form, may lead to inadequate rigidity and stiffness, hence diminishing its suitability for various applications [13].

The research and development pertaining to the modification of PBAT, utilizing both physical blending and chemical modification techniques, is experiencing a steady growth. The primary objective of modification is to enhance the tensile strength and toughness of synthesized PBAT in order to optimize its performance in various applications. Additionally, a significant aim is to minimize the cost of PBAT by using fillers, while ensuring that its biodegradability remains unaffected. Hence, via the incorporation of affordable biodegradable polymer materials, there can be a notable reduction in both the material and product costs, thereby facilitating its market adoption [14]. Starch has significant benefits when employed as a filler in PBAT composites. Starch is a polysaccharide component that rich in hydroxyl groups. It is derived from several sources and is characterized by its affordability, renewability, low-carbon footprint, and environmental sustainability [15]. Blending PBAT with starch can effectively alleviate the bottleneck issues in the use and cost of biodegradable plastics.

However, there are strong hydrogen bonding interactions between and within starch molecules, which do not exhibit thermoplastic processability. When introducing starch into a blown film based on PBAT, it is necessary to pre-plasticize the starch, breaking the inter- and intra-molecular hydrogen bonding of the starch, so that it has a flowability similar to polymer melt under a certain temperature and shear force. The starch modified by polar small molecule plasticizers is called thermoplastic starch (TPS) [16]. The PBAT film filled with starch, although not yet able to fully match the performance and cost of traditional plastics, has made significant progress in cost-effectiveness, processing, and usage performance. It has also entered mass use in specific fields such as shopping bags, garbage bags, and express delivery. However, the filling amount of starch is limited by dispersibility, high moisture absorption, and processing technology, The challenge of low starch content in the filling and relatively high cost persists.

Currently, the utilization of PBAT/TPS composite material in food packaging is limited. The main reason is the augmentation of hygroscopicity and water vapor permeability, coupled with a reduction in the inhibitory impact on microbes, subsequent to the introduction of starch [9]. In contemporary times, there exists a prevailing inclination among customers to seek food products that possess desirable flavor profiles, exhibit high standards of safety, and are devoid of preservative agents. Hence, the significance of biodegradable antimicrobial packaging materials is notably pronounced. Firstly, the utilization of antibacterial materials can contribute to the preservation of food quality and the mitigation of microbial contamination. Secondly, incorporating antibacterial agents into packaging film can effectively diminish the need for excessive preservative usage. Lastly, the implementation of biodegradable food packaging offers the advantage of convenient disposal after it has been discarded. The incorporation of active antibacterial agents into PBAT film for antibacterial packaging is a viable solution to address the growing need for enhanced packaging capabilities, and holds promising prospects for further research in the field.

The inclusion of the antibacterial agent in the biodegradable antibacterial packaging film can be achieved by two methods: direct addition into a master batch, which is subsequently incorporated into the packaging material, or application onto the surface of the packaging material following treatment. Additionally, the antibacterial agent can be blended with polymer components during processing [17]. Jiang et. al. used ginger essential oil and tea tree essential oil as antibacterial components and introduced them into PLA/PBSA composite substrate. Through blending modification, extrusion casting and other processes, antibacterial films were obtained [18]. The $O_2$ and $CO_2$ transmittance of ginger essential oil film is more suitable, and the preservation effect is optimal, which can extend the shelf life of broccoli by 2-3 days. Silva et al. introduced cellulose nanowhiskers into PBAT/TPS and PLA systems to enhance the inhibitory effect of the film against the molds, aerobic bacteria, and fungi, which is suitable for the preservation of fresh vegetables and fruits [9]. Currently, there is a rapid growth of research focused on the incorporation of antibacterial agents into food packaging materials, as well as exploring the potential synergistic effects of combining various antibacterial agents. Wongthanaroj et al. used cellulose nanocrystals or lignin nanoparticles to mix with hydroxycoumarin to increase the antibacterial activity of PLA films respectively [19]. The results showed that the film prepared by compounding 3% lignin nanoparticles with 15% hydroxycoumarin had the strongest antibacterial and free radical scavenging ability with PLA, and the barrier of PLA film increased when 3% cellulose nanocrystals or lignin nanoparticles were added to PLA. Natural antibacterial agents such as chitin, catechins, and nano lignin are usually added in excess of 10% to exhibit excellent antibacterial effects in the food packaging [20,21]. Taking chitin as an example, the powder form makes it difficult to evenly disperse in the polymer matrix when added in high amounts. Agglomeration of powder additives can easily lead to stress concentration in the film, hence causing a reduction in its mechanical properties [22]. In addition, the compatibility between natural antibacterial agents and polymer systems will also affect the mechanical and processing properties of the film.

Overall, the current biodegradable antibacterial packaging films still encounter the problem of higher prices, poor dispersion of antibacterial agents, poor strength and apparent quality, comparing to tradition packaging materials.

3. **Perspective**

The development of functional biodegradable film systems suitable for food contact and preservation is currently the mainstream work in the field. More specifically, here are several main topics: the effects of various plasticizers and their dosage on food contact migration and water absorption in PBAT and thermoplastic starch (TPS) blown film; the PBAT/PLA-based heterogeneous blend film containing natural antioxidants (preferable the plant extracts); And additives of bio-based raw materials to the PBAT/PLA system to reduces costs and endows the film with UV resistance and better barrier properties.

The industry expects to develop a series of functional biodegradable film systems that are suitable for the storage and preservation of fruits, vegetables, or snacks in the future, with continuously improved food accessibility, antibacterial properties, oxygen resistance, high mechanical strength, stable size, low moisture absorption, and various gas permeability. The studies should also examine the impact on the storage quality and anti-aging properties of model vegetables and snacks, in order to explore the key factors for extending the shelf life of vegetables, fruits, and snacks and their preservation mechanism, to provide theoretical supports for the promotion and use of biodegradable modified atmosphere packaging materials and enhancing the commercial competitiveness.

**Acknowledgments:** This work is financially supported by the Guangdong Provincial Science and Technology Program (2023A0505050146), Taizhou Municipal Science and Technology Program (22gya19), the Medical Engineering Fund of Fudan University (yg2021-005, yg2022-008), the Fudan-Yiwu Fund (FYX-23-102), and the RIZT Industrial Program (2022ZSS09, 2023CLG01, 2023CLG01PT).